\documentclass[traditabstract]{aa}    

\usepackage{txfonts}
\usepackage{longtable}
\usepackage{natbib}
\usepackage{graphicx}

\bibpunct[, ]{(}{)}{;}{a}{}{,}
\usepackage{supertabular}

\usepackage{color}

\newcommand{\he}{HE\,0338-3945}
\newcommand{\cs}{CS\,31082-001}

\newcommand{\fodd}{$f_\mathrm{odd,Ba}$}
\newcommand{\feh}{[Fe/H]}

\begin{document}

\title{The odd-isotope fractions of Barium in CEMP-r/s star HE 0338-3945 and r-II star CS 31082-001\thanks{Based on observations
carried out at the European Southern Observatory, Paranal, Chile (Proposal number 170.D-0010 and 165.N-0276(A)).}}

\author{
X. Y. Meng\inst{1},
  W.Y. Cui\inst{1,2},
J. R. Shi\inst{3},
X. H. Jiang\inst{1},
G. Zhao\inst{3},
  B. Zhang\inst{1}
 \and
  J. Li\inst{1}
}

\offprints{W.Y. Cui; \email{wycui@bao.ac.cn; wenyuancui@126.com}}
\institute{Department of Physics, Hebei Normal University,  No. 20 South 2nd Ring Road East, Shijiazhuang 050024, China
\and School of Space Science and Physics, Shandong University at Weihai, Weihai 264209, China
\and National Astronomical Observatories, Chinese Academy
of Sciences, 20A Datun Road, Beijing 100012, China\\
}

\date{Received  / Accepted }

  \abstract{
We report the first measurement of the odd-isotope fractions for barium, \fodd\, in two extremely metal-poor
stars: a CEMP-r/s star \he\ (\feh\,$=-2.42\pm0.11$) and an r-II star \cs\ (\feh\,$=-2.90\pm0.13$). 
The measured \fodd\ values are $0.23\pm0.12$ 
corresponding to $34.3\pm34.3$\% of the r-process contributions
for \he\ and $0.43\pm0.09$ corresponding to $91.4\pm25.7$\% of the r-process contribution
to Ba production for \cs. The high r-process signature of barium in \cs\ ($91.4\pm25.7\%$) suggests that the majority 
of the heavy elements in this star were synthesised via an r-process path, while the lower r-process value 
($34.3\pm34.3\%$) found in \he\ indicates that the heavy elements in this star formed through 
a mix of s-process and r-process synthesis. These conclusions are consistent with studies based
on AGB model calculations to fit their abundance distributions.
 }


\titlerunning{The odd-isotope fractions of Barium in metal-poor stars}
\authorrunning{Meng, Cui, Shi, et al.}

\maketitle
%
\section{Introduction}\label{Sect:intro}

Heavy elements  ($Z>30$) are produced mainly via two neutron-capture processes, i.e., rapid (r-)
and/or slow (s-) process. The occurrence of the s- and r-processes depends on
whether the timescale for neutron capture is slower or faster than that of $\beta$-decay processes.
The s-process occurs in low- and intermediate-mass stars
($1\leq M($M$_\odot)\leq 8$) during their asymptotic giant branch (AGB) phase, and is
supported by observational evidence and theoretical studies \citep{bus99,her05}.
Explosive conditions are usually suggested for the r-process, such as occur in core-collapse supernova
 \citep[SNe\,II, $M($M$_\odot)> 8$,][]{woo94} and neutron-star mergers \citep{ros00}.
Usually, SNe\,II is thought as a popular site for the r-process \citep[see, e.g.,][]{woo94,tak94,ter02,wan02,tho03}.
However, the latest model simulations found that SNe\,II do not produce the necessary neutron fluxes 
for r-process synthesis, whereas the neutron star merger do \citep[see][]{wan11,jan12,bur13}. 
Therefore, from a theoretical point of view, 
neutron star mergers are now the most likely scenario for r-process \citep[see Wanajo et al. 2014,][and references therein]{gor15}.

In the solar system, most of the heavy elements have contributions from both s- and r-processes,
with for instance about 82\% of Ba and 6\% of Eu being produced by the s-process, and the rest
produced by the r-process \citep{arl99}. Thus, Ba is usually
regarded as the representative element for the s-process, while Eu plays the same role for the r-process. Because the
lifetimes of massive stars are shorter than that of low- and intermediate-mass stars, the r-process 
should dominate the production of the heavy elements in the early universe.
Indeed, based on observational results, \citet{tru81} suggested that most of the heavy elements in
very metal-poor stars originate from the r-process. This proposal is supported by a
quantitative calculation using a chemical evolution model of the Galaxy \citep{tra99}. 
Likewise, \citet{mas01} found a low value of $\mathrm{[Ba/Eu]}\sim-0.7$ in metal-poor stars, 
and they suggested that Ba is produced by only the r-process in the early history
of the Galaxy.

Carbon-enhanced metal-poor (hereafter CEMP) stars are a kind of stars with $\mathrm{[C/Fe]}>+1.0$ and $\mathrm{[Fe/H]}<-2.0$, 
which is divided into four sub-classes: CEMP-s, CEMP-r, CEMP-r/s, CEMP-no stars based on the abundance pattern of neutron-capture 
elements \citep{bee05}. The enhanced carbon and s-process materials for CEMP-s stars are supposed to be transferred from its massive companion 
during the AGB phase via either Roche Lobe overflow or wind accretion in a binary system. The binarity has been confirmed through 
radial-velocity monitoring for most of them \citep{luc05}. CS 22892-052 is the only CEMP-r star found up to now \citep{sne03}, 
which is also belong to the r-II group \citep{bee05}. It is usually supposed that the r-II stars formed from a molecular cloud that has been 
polluted with r-enriched materials \citep[][and references therein]{sne08}.

CEMP-r/s stars exhibit large over-abundances for both the s-element Ba and the r-element Eu ($\mathrm{[Eu/Fe]}>1$
and $0<\mathrm{[Ba/Eu]}\leq0.5$), as first noted by \citet{bar97} and \citet{hil00}.
The origin of their peculiar abundance pattern  is very puzzling, as the r- and s-processes need very different astrophysical conditions.
Many scenarios have been proposed to explain the abundance peculiarities in CEMP-r/s stars,
but none can coherently interpret all the observational properties \citep[see details in][and references therein]{jon06}.
A popular explanation is the double pollution mechanism, i.e., for a CEMP-r/s star the C and s-process meterials come from its 
AGB companions similar to CEMP-s stars and the binary system formed from a molecular cloud which had already been 
polluted with r-enriched materials. Based on such a scenario, a series of theoretical calculations
has been carried out, and the observed abundance patterns of heavy elements
in CEMP-r/s stars have been fitted well \citep{bis12,cui10,cui13a,cui13b,cui14}.
A much stronger basis for this scenario is, however, desirable and the unambiguous 
detection of r-process isotopes and s-process isotopes is important for evaluating this picture.
For CEMP stars lying in the range of \feh\,$<-3.4$ \citep{aok07}, however, their abundance patterns 
suggest that these stars formed with this high carbon abundance from (possible pop III) ISM enrichment \citep{spi13}.
In fact, a lot of CEMP-no stars lie in this region \citep[see][]{mas10}.

There is a direct way to detect r- and s-process signatures in a star by measuring
\fodd\ from the Ba\,II resonance line profile
(4554\,{\AA}), using a method first suggested by
\citet{cow89} and \citet{mag93}. Barium has five stable isotopes, i.e., $^{134}$Ba,
$^{135}$Ba, $^{136}$Ba, $^{137}$Ba, $^{138}$Ba, and they are all produced
through neutron-capture processes. The two even isotopes, $^{134}$Ba and $^{136}$Ba, are however shielded by
$^{134}$Xe and $^{136}$Xe on the r-process path, and thus they can only be produced by the s-process.
Therefore, for a fixed barium abundance, we can say that different \fodd\ values correspond 
to different r-process contributions to the barium production, with higher \fodd\
values corresponding to higher r-process contributions.
Based on the solar abundances of \citet{arl99}, the \fodd\
value is 0.46 for the pure-r-process production of barium, while it is only 0.11 for the pure-s-process. 
In other words, a measured \fodd\ value of approximately 0.46, implies a pure-r-process 
origin of barium, and on the contrary a value of 0.11 means a pure-s-process origin for the barium.
Generally, \fodd\ values lie between 0.11 and 0.46, which means that the
barium is produced by both s- and r-processes.
Values of \fodd\ can be measured by fitting the profile of the strong Ba\,II resonance line at 4554\,{\AA},
because this line usually experiences significant hyperfine splitting for the odd isotope contributions, which leads to the line profile being asymmetric.

Many attempts have been made to obtain
\fodd\ in the metal-poor subgiant HD\,140283 \citep{mag95,lam02,col09,gal10,gal12}.
However, conflicting conclusions have been obtained based on the apparently
different \fodd\ values obtained. Small \fodd\ values support an s-process dominated production of barium
\citep{mag95,gal10,gal12}, while large values support an r-process origin of barium for
the metal-poor star HD\,140283 \citep{lam02,col09}. One reason for the discrepancy is that the fractions obtained
in the above works were all measured under the assumption of local thermodynamic equilibrium (LTE),
which introduces large uncertainties in the results in this case, because
the profile of the resonance line in metal-poor stars suffers strong non-LTE (NLTE)
effects \citep{mas99, sho06}. Another reason is that HD\, 140283 has a low barium abundance
\citep[$\mathrm{[Ba/H]}<-3.1$, see][and reference there in]{gal10}, and thus, the resonance line may be too weak to
get a reliable value of \fodd. In other hand, \citet{col09} discussed the three-dimensional (3D) effects 
to the asymmetry of the barium line. \citet{gal15} have re-evaluated \fodd\ with 3D atmosphere model for HD 140283 
and found that their new result suggests a stronger r-process signature in this star 
than any other study before. They also examined the effects of using 1D LTE synthesis 
to measure \fodd\ in the 4554 line by fitting the 1D profiles to a 3D s- and 3D r-process 4554 line, 
and find that the 1D synthesis will invariably favour an s-process signature in both cases. 
Based on the measured \fodd\ values in a
sample of disk stars, \citet{mas06} reported higher r-process contributions in thick-disk
stars compared to those in thin-disk stars. However, there is no such study to determine the values of \fodd\ 
in metal-poor stars enriched in neutron-capture elements.

Although studies are currently focused mainly on the element abundance patterns,
isotope abundance ratios for heavy elements will likely be able to give more critical constraints on the
theoretical models. In this paper, we report the Ba odd-isotope fractions for two stars rich in 
neutron-capture elements, \he\, and \cs. The paper is organized as follows. Section 2 presents
the observations and data reduction, and barium isotope ratios analysis is presented in Section 3. 
The determined isotope fractions of barium appear
in Section 4 and the r/s contributions to barium production for \he\ and \cs\
are discussed in Section 5. Conclusions are presented in Section 6.

\section{Observations and Data Reduction}\label{Sect:data}

Two metal-poor and neutron-capture element enriched stars, \he\, and \cs, have been 
observed with the VLT spectrograph UVES. 
\he\, was observed with a
resolution of $\sim 30\,000-40\,000$ on the nights of 11 December and 23 to 25, 2002.
For the spectrum including the Ba\,II line at 4554\,{\AA}, the total exposure times were 9.5\,h,
while 6\,h of exposure for the region with Ba\,II lines at 5853 and 6496\,{\AA}.
The individual exposure times ranged from 30 to 75\,min \citep[see][for details]{jon06}.
While, the \cs\, was observed during
October 2000. A high resolution of $R=75\,000$ 
was achieved for the spectrum in the Ba\,II line (4554, 5853 and 6496\,{\AA}) region. 
The total exposure times are two and three hours for setting 380-510\,nm 
and 480-680\,nm, respectively \citep[see][for details]{hil02}.
The raw data were downloaded from 
the European Southern Observatory (ESO) archive. 

The spectra were reduced with the program
designed originally for the FOCES spectrograph \citep{pfe98}, which has been modified to work for UVES spectrograph.
The program works under the IDL environment. Cosmic rays and bad pixels were
removed by careful comparisons of the exposures for the same objects. The instrumental 
response and background scatter light were also considered during the data reducing.
It is worth noting that we have not found the proper lamp exposures with the same settings for the red region 
taken by the CD\#3 equipped on the red arm of the UVES spectrograph for both of our program stars. 
Thus, the pipeline-reduced spectra have been adopted for the red 
spectra region ($>$\,500\,nm).

\section{Barium Isotope Ratios analysis}
\subsection{Stellar Parameters}

As \citet{aok03} pointed out, the isotope ratios are insensitive to the stellar atmospheric
parameters, we directly adopted the stellar parameters derived by \citet{jon06} and \citet{hil02} for both of 
our program stars, the detail information is listed in Table\,\ref{stellardata}. For \he, \citet{jon06} found that the values of effective temperature ($T_\mathrm{eff}$)
from photometry such as $B-V$, $V-R$ and $V-I$ ranges between 6100\,K and 6350\,K. 
The final adopted effective temperature was obtained through analysis of the Balmer H$\beta$ and H$\delta$
lines, while the adopted $\mathrm{log}\,g$, \feh, and $V_\mathrm{mic}$ were determined
spectroscopically using the optimization routine
of \citet{bar05}, which is based mainly on the analysis of the weak Fe and Ti lines.

\citet{hil02} computed the effective temperature for \cs\ from multicolor information 
using the \citet{alo99} color-temperature transformations. They found that the values of 
$T_\mathrm{eff}$ from $B-V$, $b-y$, $V-R_\mathrm{C}$, $V-I_\mathrm{C}$ 
(subscript C indicating the Cousins system) and $V-K$ ranges from 4818\,K
to 4917\,K (details see their table\,2). 
$T_\mathrm{eff}=4825\pm100\,K$ was adopted finally, which was derived
from multicolor index using the \citet{alo99} color-temperature transformations,
and it is consistent with that obtained from the excitation equilibrium of Fe\,I lines. The surface gravity
was derived with the ionization equilibrium of Fe and Ti lines, and
the microturbulence velocity was obtained by requiring the strong and weak Fe lines to give the same abundances.
Detailed information regarding the stellar parameters and the error bars are shown for both stars in Table~\ref{stellardata}.

\begin{table*}
\begin{center}
\caption{Stellar parameters and Ba abundances of \he\ and \cs\ adopted from
\citet{jon06} and \citet{hil02}, respectively.}
\label{stellardata}
\begin{tabular}{lccccccc}
\hline
\hline
Name & $V$ & $S/N$ &$T_\mathrm{eff}$ & $\log g$ & $V_{mic}$ & \feh\ & [Ba/H]   \\
& (mag) &&(K) & (dex)& (km\,s$^{-1}$) & \\
\hline
\he        &$15.333\pm0.007$ &$>70$& $6160\pm100$  & $4.13\pm0.33$ & $1.13\pm0.22$ & $-2.42\pm0.11$ & $-0.01$ \\
\cs        & $11.674\pm0.009$&$>250$& $4825\pm100$ & $1.5\pm0.3$ & $1.8\pm0.2$ & $-2.90\pm0.13$ & $-1.73$ \\
\hline
\end{tabular}
\end{center}
\end{table*}

\subsection{Analysis Method}

In order to study the characteristics of the neutron-capture processes in stars enriched in n-capture elements, 
and provide critical constraints on AGB models,
it is important to determine the fraction of the odd Ba isotopes, \fodd. In this work,
\fodd\ is defined as
\begin{small}
\begin{equation}\small
f_\mathrm{odd,Ba}=\frac{N(^{135}\mathrm{Ba})+N(^{137}\mathrm{Ba})}{N(^{134}\mathrm{Ba})+N(^{135}\mathrm{Ba})+N(^{136}\mathrm{Ba})+N(^{137}\mathrm{Ba})+N(^{138}\mathrm{Ba})}.
\end{equation}
\end{small}
The significant hyperfine splitting for the odd isotopes leads to the profiles of the Ba\,II resonance
line at 4554\,\AA\ being asymmetric, while it has negligible effect on the
subordinate line profiles, such as the Ba\,II 5853 and 6496\,{\AA} lines. 
An additional property that contributes to the asymmetry of spectral features, including barium, is stellar convection.

Following the approach adopted by \citet{mas06}, we first derived the Ba abundance from 
the subordinate lines at 6496 and 5853\,{\AA}, and then 
used \fodd\ as a free parameter to fit the profile of the Ba\,II resonance line at 4554\,{\AA} subject to
the fixed Ba abundance derived from the above two Ba\,II subordinate lines.
The opacity sampling MAFAGS model atmospheres from \citet{gru04} and \citet{gru09}, 
and the Ba atomic model from \citet{mas99} and \citet{mas06} were adopted.
The IDL/Fortran SIU software package of \citet{ree91} was used to compute the
synthetic line profiles. 

In order to find the best fit from a set of synthetic spectra to the observed ones, 
we calculated the values of reduced $\chi^2$, $\chi_\mathrm{r}^2$.\footnote{
\begin{eqnarray*}
\chi_r^2=\frac{1}{\nu-1}\sum_{i=1}\frac{(O_i-S_i)^2}{\sigma_i^2},
\end{eqnarray*}
where $O_i$ is the observed continuum-normalised flux, $S_i$ are the 
synthetic spectral points, $\nu$ is the number of degrees of freedom in 
the fit, and $\sigma_i$ is the 
standard deviation of the data points defining the continuum of the observed spectrum \citep{smi98}. 
$\sigma$ is defined as $\sigma=(S/N)^{-1}$, where $S/N$ is measured in roughly 1\,\AA\ interval on 
either side of the spectral line referred.} 
When measuring the \fodd\ values, it generally requires three parameters: 
the wavelength shift, a continuum level
shift to match the synthetic continuum, as well as  macroturbulence to make a comparison between the 
observed and synthetic spectra. In this work, the continuum level is fixed after a careful renormalisation of the observed continuum over
a window of each Ba\,II line at $\lambda$\,4554\,\AA\ \citep[as done in][]{gal10}. 
The number of points in the line profile at $\lambda$\,4554\,\AA\ used in
our fitting are 23 and 22\, pixels for \he\ and \cs, respectively. The total degrees of freedom $\nu$ were thus 
23 or 22 minus two fitting parameters for the above two stars, respectively. For the best fit case, the 
$\chi_\mathrm{r}^2$ value is expected to be the minimum one. The instrumental broadening was derived from a 
Th-Ar lamp spectrum with a Gaussian fit, and the Th-Ar lamp spectrum
was taken with the same instrumentation setup as the object exposures. As discussed by \citet{smi98},
any non-Gaussian extended wings of the instrumental profile are weak, and have no measurable effect 
on the line-profile fitting. In our synthetic profiles the broadening due to rotation is also involved, and
the projected rotational velocity $v\,\mathrm{sin}\,i=1.5$\,km\,s$^{-1}$ is adopted for both \he\ and \cs. 
Here, $v$ is the surface equatorial rotational velocity.
This is reasonable because \citet{smi98}
have shown that $v\,\mathrm{sin}\,i$ is less than 3\,km\,s$^{-1}$ for old stars.

\section{Results}
\subsection{\he}
As mentioned above, we firstly determined the barium abundance 
($\mathrm{[Ba/Fe]}=2.47\pm0.11$) in \he\ using the two subordinate lines, 
and included the influence of NLTE effects.
The NLTE barium abundances determined using the subordinate lines at 6496 and 5853\,\AA\ are
listed in Table~\ref{ba.abundance}, where the LTE Ba abundances are also included for comparison.
The total NLTE barium abundance in \he\ are shown in Table~\ref{nltefodd}.
From Table~\ref{ba.abundance}, it can be seen
that the NLTE corrections for both of subordinate lines are small, which is consistent with
the prediction of \citet{mas99} for the two Ba\,II lines in metal-poor
(\feh\ $<-2$) stars. Our measured Ba abundance, however, is slightly higher
than that of the LTE value (2.41) measured by \citet{jon06}. We note that
\citet{jon06} have used two weak lines at
4166 and 4524\,\AA\ and a strong resonance line at 4554\,\AA\
to determine the total Ba abundance of \he. Although the NLTE effects are small for the two weak lines, it is
large for the strong Ba\,II resonance line \citep[about $+0.2$\,dex, see][]{asp05}.
If the NLTE corrections are included, the result of \citet{jon06} is consistent with ours within errors.
The measured equivalent widths (EW) of the resonance line at 4554\,\AA\  and those of
the two subordinate lines at 6496 and 5853\,{\AA} are shown in Table~\ref{ba.abundance}. 

\begin{table}
\begin{center}
\caption{Line data, equivalent widths (EW), and barium abundances in \he\ and \cs\ obtained using several Ba lines.}
\label{ba.abundance}
\begin{tabular}{lccccc}
\hline
\hline
$\lambda$ & $\chi_{ex}$ & $\log gf$ &EW& [Ba/Fe] & [Ba/Fe]  \\
(\AA) &(eV) & &(m\AA)&LTE&NLTE\\
\hline
\he   \\
\hline
5853.668   & 0.604 & -1.000 &83.7& 2.44 & 2.45 \\
6496.897   & 0.604 & -0.377 &113.4& 2.49 & 2.49  \\
4554.029  & 0.000 & 0.170  &187.7 &--&--\\
\hline
\cs \\
\hline
5853.668   & 0.604 & -1.000 &78.5& 1.14 & 1.15 \\
6496.897   & 0.604 & -0.377 &115.1& 1.12 & 1.13  \\
4554.029  & 0.000 & 0.170  &181.5 &--&--\\
\hline
\vspace{-11mm}
\end{tabular}
\end{center}
\end{table}

\begin{figure*}[bhtp]
\begin{center}
  \resizebox{200mm}{!}{\includegraphics{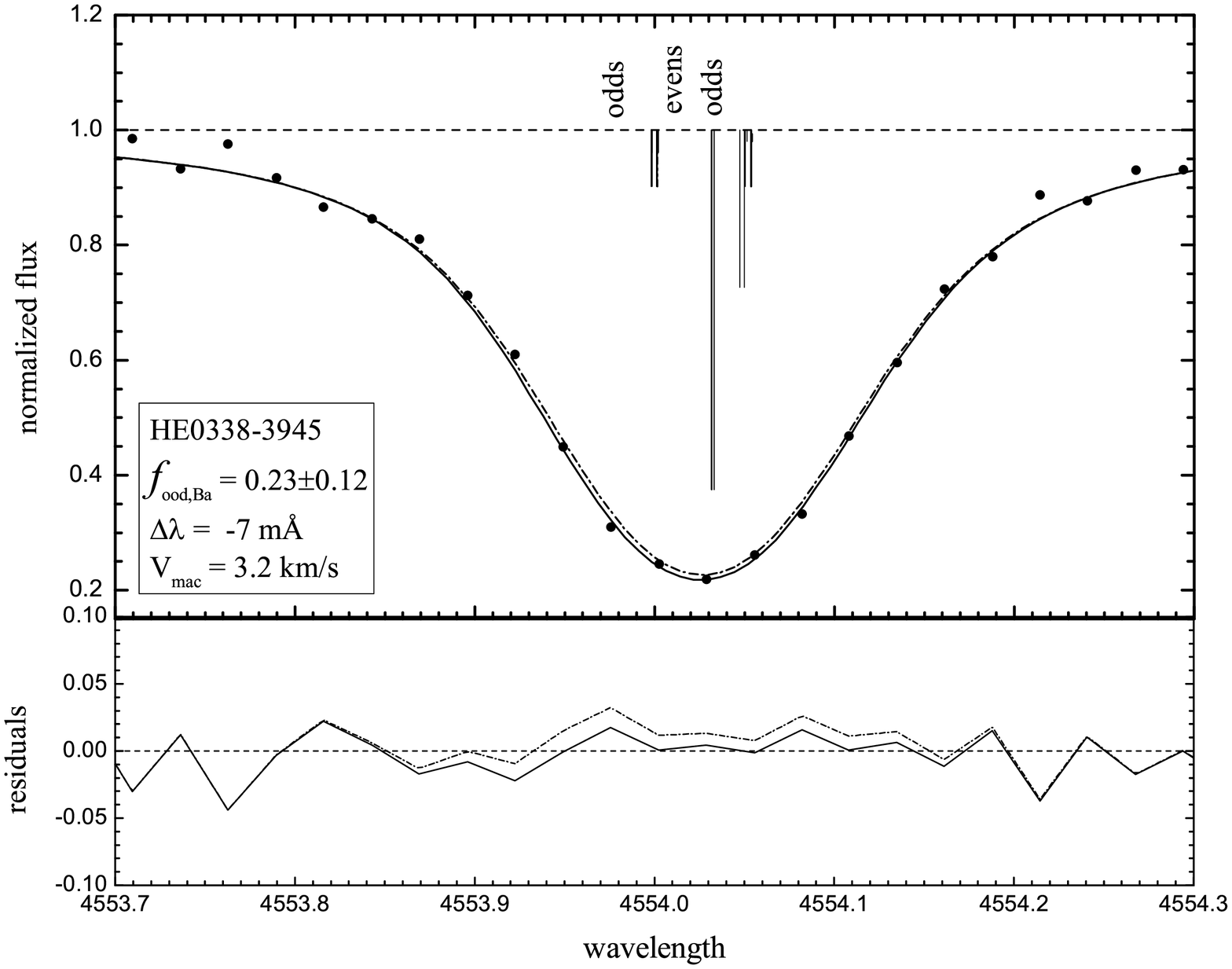}\hspace{-35mm}\includegraphics{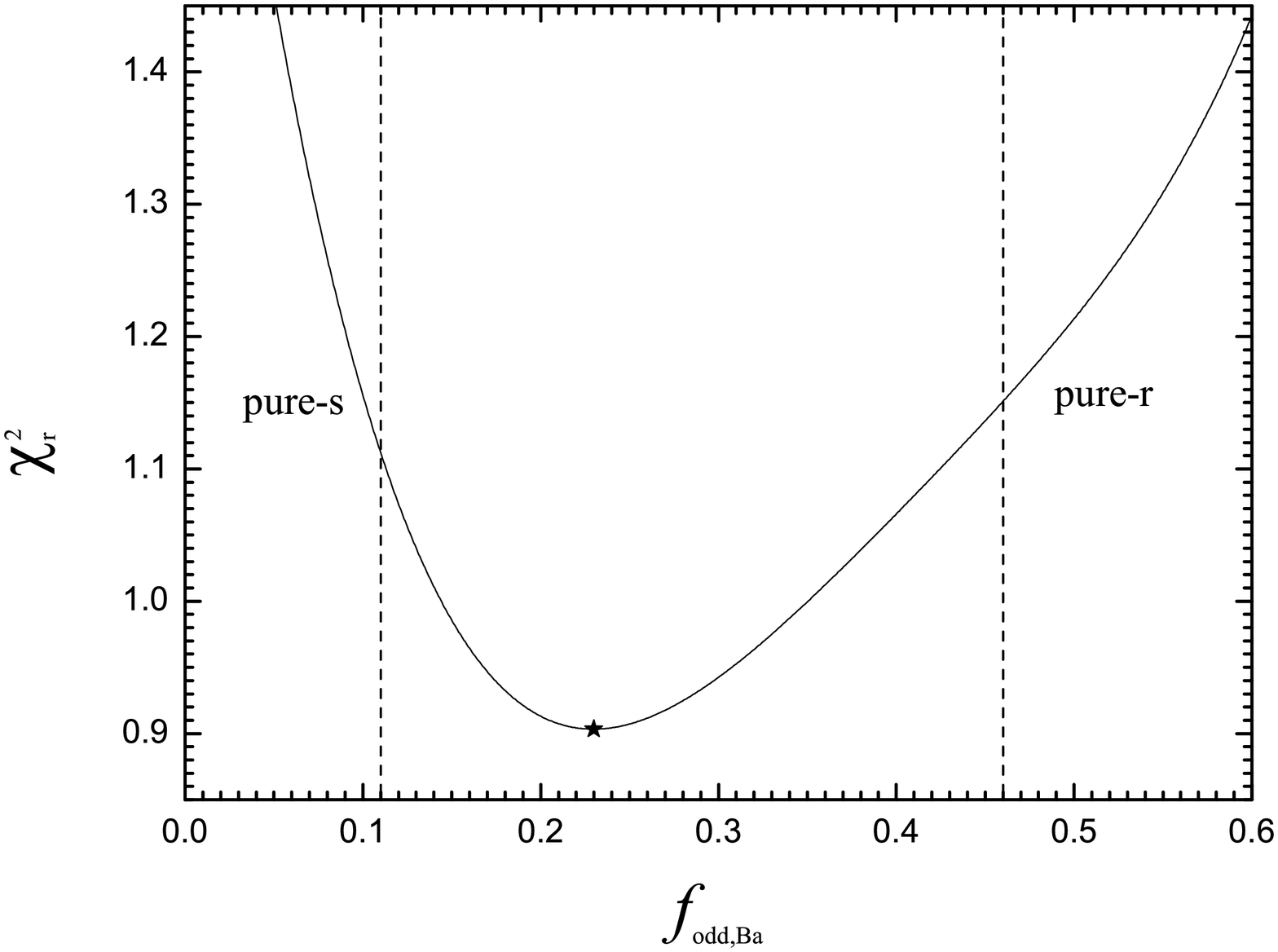}}
 \hspace{-2mm}
  \caption{Left panel: the best statistical fit synthetic profile obtained with \fodd\,$=0.23$ and 
NLTE line shapes for the observed (filled circles) Ba\,II resonance line at 4554\,\AA\ in \he\ 
  with the residual plots below. For comparison, a line with \fodd\,$=0.11$ (i.e. $0.23-\sigma$) and residual
  have been plotted (dash-dot line). The value for $V_\mathrm{mac}$ has been optimised to one that 
  minimises $\chi_\mathrm{r}^2$, and the value for [Ba/Fe] remains the same. 
  Right panel: we show the $\chi_\mathrm{r}^2$ fit for the 4554\,\AA\ line, the star
  shows where the minimum of the fit lies.\label{fitting4554}}
\end{center}
\end{figure*}

\begin{figure*}[bhtp]
\begin{center}
\resizebox{200mm}{!}{\includegraphics{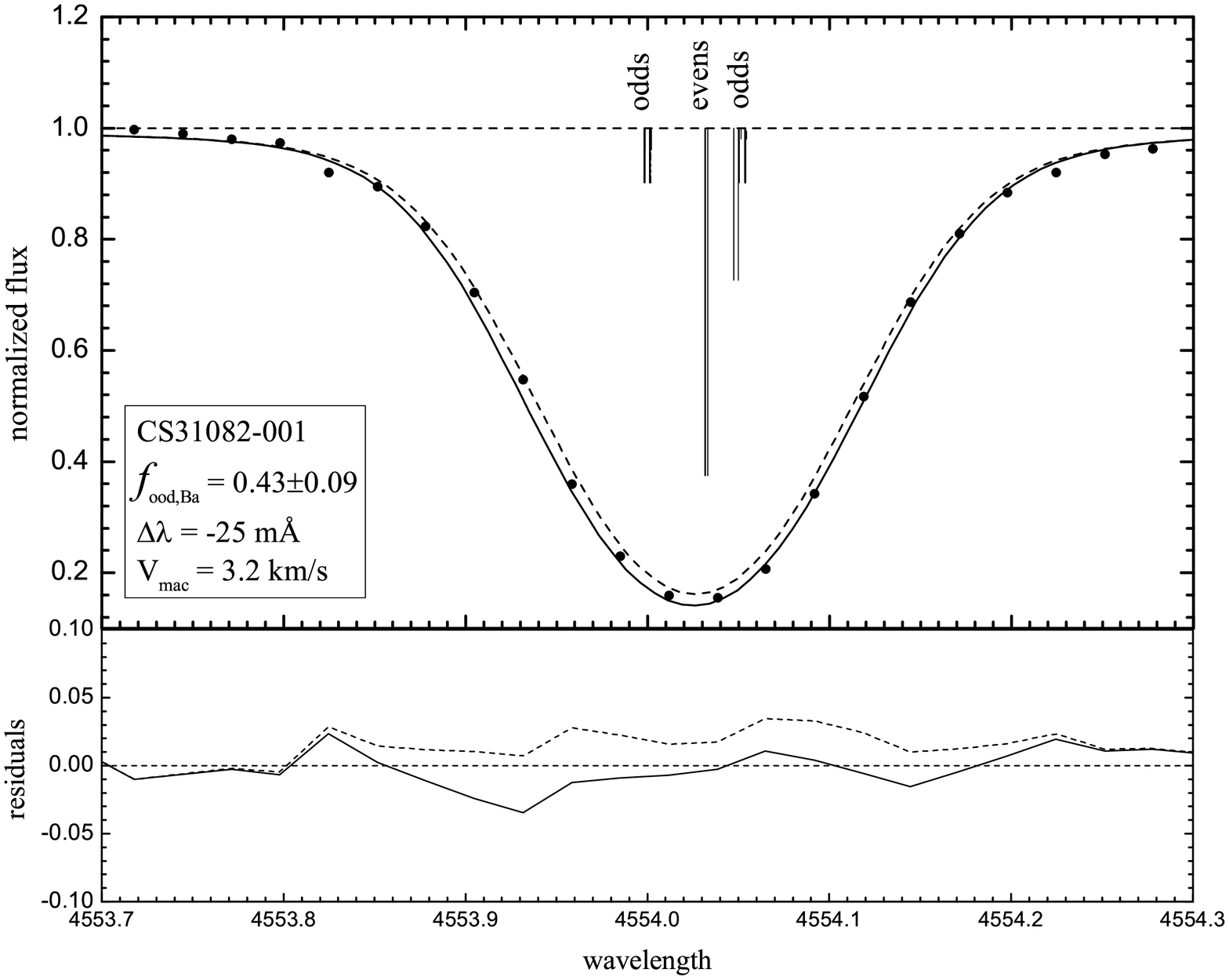}\hspace{-35mm}\includegraphics{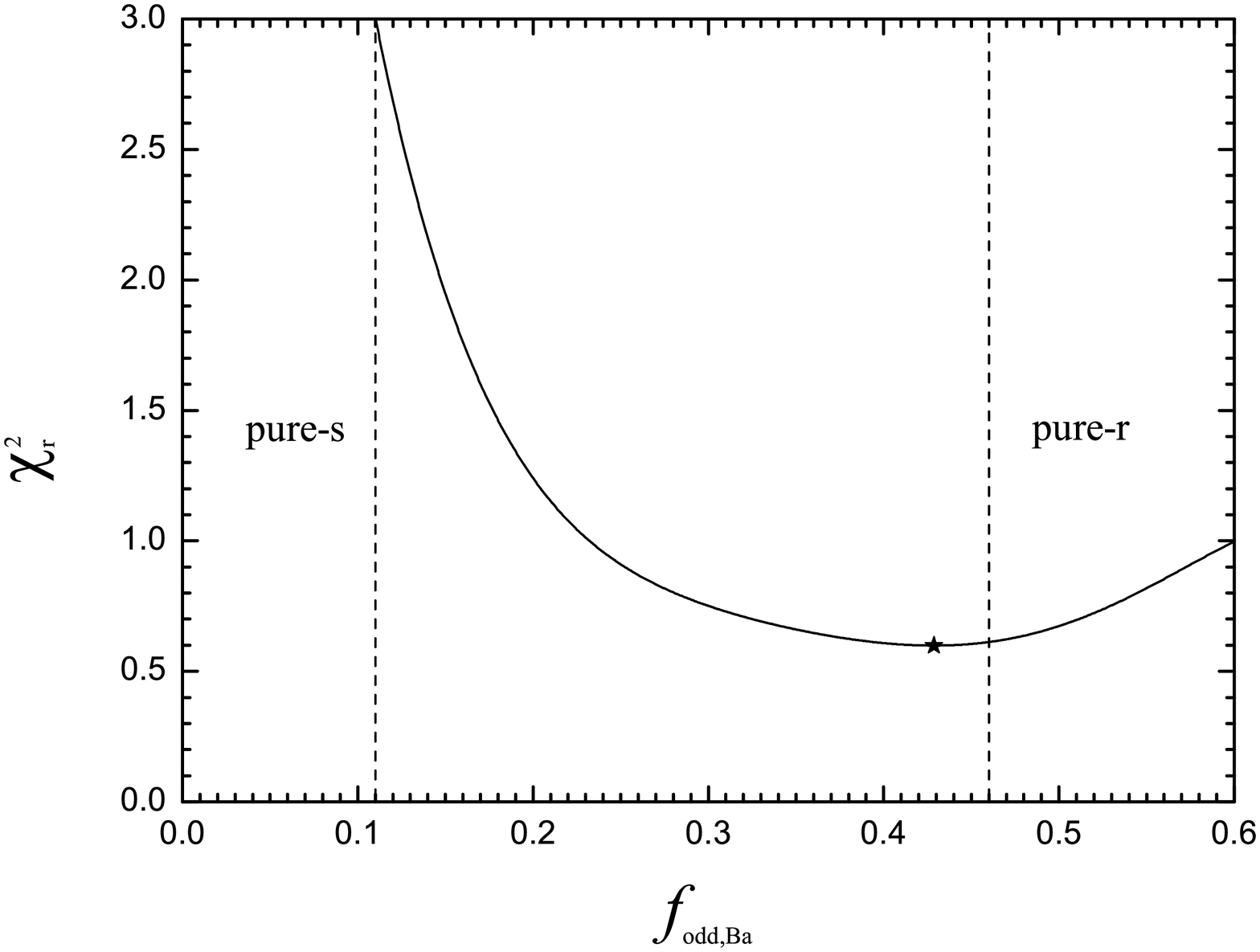}}
 \hspace{-8mm}
  \vspace{0mm}
  \caption{Left panel: the best statistical fit synthetic profile obtained with \fodd\,$=0.43$ and 
NLTE line shapes for the observed (filled circles) Ba\,II resonance line at 4554\,\AA\ in \cs\ 
  with the residual plots below. For comparison, a line with \fodd\,$=0.16$ (i.e. $0.43-3\sigma$) and residual
  have been plotted (dash-dot line). The value for $V_\mathrm{mac}$ has been optimised to one that 
  minimises $\chi_\mathrm{r}^2$, and the value for [Ba/Fe] remains the same. 
  Right panel: we show the $\chi_\mathrm{r}^2$ fit for the 4554\,\AA\ line, the star
  shows where the minimum of the fit lies. \label{fitting}}
\end{center}
\end{figure*}

\begin{table*}
\begin{center}
\caption{NLTE Ba abundances and Ba odd-isotope fractions for \he\ and \cs, respectively.}
\label{nltefodd}
\begin{tabular}{lcccc}
\hline
\hline
Name &[Ba/Fe]  & $\sigma$(total) & \fodd & $\Delta$ \\
\hline
\he        & $2.47$&$0.11$ & $0.23$ & 0.12\\
\cs        & $1.14$ &$0.05$ & $0.43$ & 0.09\\
\hline
\end{tabular}
\end{center}
\end{table*}

Figure\,\ref{fitting4554} shows the process for determining the \fodd\ value through
fitting the profile of the Ba\,II resonance line at 4554\,\AA\ for \he. Considering the strong
dependence of the \fodd\ value on the Ba abundance, and the larger NLTE effect
for the Ba\,II resonance line, the NLTE correction has also been included in order to get a reliable \fodd\ value.
In the current work, only a spectrum with $S/N>70$ obtained in a single exposure, was used
in order to avoid uncertainties in \fodd\ due to possible
changes in the resonance line profile at 4554\,{\AA} caused by the co-adding process. 
As the high Ba abundance in \he,
the resonance line is strong enough to ensure the accuracy of \fodd.

As discussed by \citet{mas06} we have no arguments to fix the macroturbulence value, 
it was allowed to be free during the analysis process. 
The macroturbulence value of the Ba\,II line at 4554\,\AA\ was found to be $V_\mathrm{mac} = 3.2$\,km\,s$^{-1}$,
and a small wavelength shift, $\Delta\lambda=-7$\,m\AA, is also required by the $\chi_\mathrm{r}^2$ fit. 
The best statistical fit for the 4554\,\AA\ line and residual (synthetic $-$ observed profile) are illustrated 
in the left panel of Figure\,\ref{fitting4554}, where the total NLTE Ba abundance,
$\mathrm{[Ba/Fe]}=2.47$, and \fodd\,=$0.23$ were used. 
The synthetic profile for this barium line with \fodd\,$=0.11$ (i.e. $0.23-\sigma$) is also been plotted for comparison. 
It can be seen in the residual plots for the Ba\,II 4554 line that the fits of the synthetic line with $\sigma$ deviation from the 
\fodd\,$=0.23$ are very poor.

In the right panel of Figure~\ref{fitting4554}, we showed the $\chi_\mathrm{r}^2$ versus \fodd. 
It can be seen that the minimum 0.903 of $\chi_\mathrm{r}^2$ is obtained at \fodd\, $=0.23$ where the gradient of the 
$\chi_\mathrm{r}^2$ curve is zero. We noted that
the spectrum of \he, used for the isotope fraction determination, has a slightly low S/N ratio (S/N\,$>70$).
However, \he\ has a sufficiently large barium abundance ($\mathrm{[Ba/H]}=0.05$), 
which is much higher than the low limit $\mathrm{[Ba/H]}=-2.0$ suggested by \citet{mas06} for such a method. 
This means that both the
subordinate barium lines at 6496 and 5853\,{\AA} and the resonance line at 4554\,\AA\ in \he\
are strong enough to obtain an accurate \fodd\ (see Table~\ref{ba.abundance} and \ref{nltefodd}).
Through a test, \citet{mas06} found that their method failed to give a reliable \fodd\ value at $\mathrm{[Ba/H]}<-2$, because the
resonance line is weak and less sensitive to the variations of \fodd\ in these cases.
Therefore, we are can obtain a 
reliable \fodd\ value of $=0.23\pm0.12$ for \he\ using the above method.

\subsection{\cs}

The barium abundance in \cs\ is $\mathrm{[Ba/Fe]}=1.14\pm0.05$, which was also derived from the
two subordinate lines (6496 and 5853\,\AA). 
Our result is slightly lower ($-0.03$ dex) than the LTE value of
\citet{hil02}. The reason is that although the
Ba\,II resonance line has also been used, 
\citet{hil02} also used six additional weak lines to determine the Ba abundance,
and the weak lines do not suffer large NLTE effects \citep{mas99,asp05}.
Their final Ba abundance is a average of the abundances derived from every barium line used,
and thus they made the LTE Ba abundance very close to the NLTE value.
Both the NLTE and LTE barium abundances are shown 
in Table~\ref{ba.abundance}. In addition, the total NLTE barium abundance are listed in Table~\ref{nltefodd}. As predicted by \citet{mas99} for metal-poor stars, it can be seen
that from Table~\ref{ba.abundance} the NLTE corrections for the two weak subordinate lines are small.
The EWs of both the resonance line at 4554\,\AA\ and the two subordinate lines at 6496 and 5853\,{\AA} are shown in Table~\ref{ba.abundance}. 

Similar to the case of \he, in \cs, the Ba-odd-isotope fraction, \fodd\, $=0.43\pm0.09$,
 was also obtained from fitting the Ba\,II resonance
line profile at 4554\,\AA.  This is a metal-poor
($\mathrm{[Fe/H]}=-2.90$) giant star with an enhanced barium abundance
($\mathrm{[Ba/Fe]}=1.14\pm0.05$) corresponding to $\mathrm{[Ba/H]}=-1.76$,
which is still higher than that of the lower limit ($\mathrm{[Ba/H]}=-2.0$) 
suggested for this method by \citet{mas06}.
Like in \he, both the two subordinate barium lines at 6496 and 5853\,{\AA} 
and the resonance line at 4554\,\AA\ are also strong enough (see Table~\ref{ba.abundance}, \ref{nltefodd}).
In addition, the signal-to-noise at the Ba\,II resonance line 4554\,\AA\ is higher than 250, thus, a reliable \fodd\ value
has been derived for \cs.

The best-fit result
for the observed profile of the Ba\,II resonance line at 4554\,\AA\ and residual are shown
in the left panel of Figure~\ref{fitting}, while the right panel shows $\chi_\mathrm{r}^2$ versus \fodd.
For the same reason presented in section 4.1, we allowed the value of macroturbulence to be a free parameter during 
our analysis process. The macroturbulence value of the Ba\,II 4554 line was found to be $V_\mathrm{mac} = 3.2$\,km\,s$^{-1}$,
and a small wavelength shift, $\Delta\lambda=-25$\,m\AA, is required when doing the $\chi_\mathrm{r}^2$ fit. 
The synthetic profile for this barium line with \fodd\,$=0.16$ (i.e. $0.43-3\sigma$) is also been plotted for comparison. 
It can be seen in the residual plots for the Ba\,II 4554 line that the fits of the synthetic line with $3\sigma$ deviation from the 
\fodd\,$=0.43$ are very poor.
The best fit has the minimum $\chi_\mathrm{r}^2$ value of 0.599
with \fodd\,$=0.43$.

\subsection{Uncertainty of the Ba-odd-isotope Fraction}

The total uncertainty in \fodd\ includes random and systematic errors \citep{mas06}.
Random errors are mainly caused by the error in the barium abundance and 
uncertainties in the stellar parameters including $T_\mathrm{eff}$, log\,$g$ 
 and the microturbulence velocity $V_\mathrm{mic}$.
The uncertainties in the atomic parameters ($\mathrm{log}gf$, $\mathrm{log}C_6$) 
of the Ba\,II resonance line at 4554\,{\AA} result in systematic
errors. A test was carried out for the possible changes in \fodd\ caused by an assumed 
variation of one item of the atomic data or stellar parameters with the other parameters fixed.
\citet{mas06} gave a detailed discussion of this subject, but only for stars
with metallicities close to the solar value. Table~\ref{uncertainty} summarizes the 
various sources of uncertainties influencing the derived odd-isotope fractions of Ba
in \he\ and \cs.

As both our sample stars have high barium abundances, the subordinate lines 
are strong enough, thus the   
uncertainty of the van der Waals damping constant is an 
important error source of Ba abundance.
A variation of 0.1\,dex in log\,$C_6$ for the subordinate lines translates 
to the 0.02 and -0.01\,dex variation in $\mathrm{log}\epsilon_\mathrm{Ba}$
and then translates to a variation of 0.014 and -0.001 in 
\fodd\ for \he\ and \cs, respectively.
The stellar parameters of \he\ and \cs\ are adopted from \citet{jon06} and \citet{hil02}, respectively. 
The errors of $T_\mathrm{eff}$, log\,$g$ and $V_\mathrm{mic}$ are $\pm$100\,K, $\pm$0.33
and $\pm$0.22 for \he, and $\pm$100\,K, $\pm$0.3
and $\pm$0.2 for \cs. Uncertainties of these stellar parameters combining  
with the difference of Ba abundance between two subordinate lines lead to the obtained
errors 0.11 and 0.05 in [Ba/Fe] for \he\ and \cs, respectively (see Table~\ref{nltefodd}).
If we assume variations of 100\,K in $T_\mathrm{eff}$ for the above two stars,
we get uncertainties in the derived \fodd\ of -0.021 and 0.086, respectively.
Variations of -0.22 and -0.2 in $V_\mathrm{mic}$ translate into the uncertainties 
in the obtained \fodd\ of -0.056 and -0.02 for \he\
and \cs, respectively. Because the macroturbulence was allowed to be a free parameter during the 
process of measuring \fodd\ values for both stars, a test for the possible uncertainties due to $V_\mathrm{mac}$
has been done. A variation of -0.2 in $V_\mathrm{mac}$ translates into the  
uncertainties in the obtained \fodd\ of -0.006 and 0.004 for \he\ and \cs, respectively.
The random errors of the obtained fraction resulting from the uncertainties of Ba 
abundance, $T_\mathrm{eff}$, log\,$g$ and $V_\mathrm{mic}$, in total,
are 0.07 and 0.09 for \he\ and \cs, respectively. The systematic effects, 
although they depend only weakly on [Ba/H] \citep{mas06}, are strong,
because a variation of 0.1\,dex in log\,$C_6$ for the Ba\,II resonance line 
at 4554\,{\AA} produces uncertainties of 0.08 and 0.01 in the values of \fodd\ obtained 
for \he\ and \cs, respectively. 

\begin{table}[hbtp]
\begin{center}
\caption{Effects on the values of \fodd\ resulting from uncertainties of atomic data and stellar parameters.}
\label{uncertainty}
\begin{tabular}{lcccc}
\hline
\hline
Input & Input & HE\,0338&Input&CS\,31082 \\
parameter&error &-3945 &error&-001\\
\hline
log\,$C_6(5d-6p)$   & +0.1 & +0.014 & +0.1 & -0.001 \\
$\mathrm{[Fe/H]}$ & -0.11 & -0.015 & -0.13 & +0.01  \\
$T_\mathrm{eff} $(K)  & +100 & -0.021 &+100 & +0.086 \\
log\,$g$  & -0.33 & -0.027 & -0.30 & -0.009  \\
$V_\mathrm{mic}$  & -0.22 & -0.056 & -0.2 & -0.02  \\
log$gf$(4554) & -0.1 & +0.047 & -0.1 & +0.016 \\
log\,$C_6$(4554)   & +0.1 & +0.084 & +0.1 & +0.01 \\
$V_\mathrm{mac}$& -0.2 & -0.006  & -0.2 & +0.004 \\
$\Delta$(total) & & $\pm0.12$ & & $\pm0.09$ \\
\hline
\vspace{-10mm}
\end{tabular}
\end{center}
\end{table}

The total errors in \fodd\ for \he\ and \cs\ are listed in Table~\ref{nltefodd}. They included the
effects of the uncertainties in the stellar and atomic parameters. The
uncertainties in the values of \fodd\ for \he\ and \cs\ are
estimated to be $0.12$ for \he, and $0.09$ for \cs,
respectively, where the uncertainty ($\pm0.01$\,dex, depending on
the $\chi_\mathrm{r}^2$ value) from fitting the $\lambda$ 4554 line profile has
also been considered. The resultant values of \fodd, with their
errors, are presented in
Table~\ref{nltefodd} for our two stars.

\section{The r/s-process Contributions}

\begin{figure}[htbp]
\begin{center}
  \resizebox{85mm}{!}
{ \vspace{-8mm}
\includegraphics{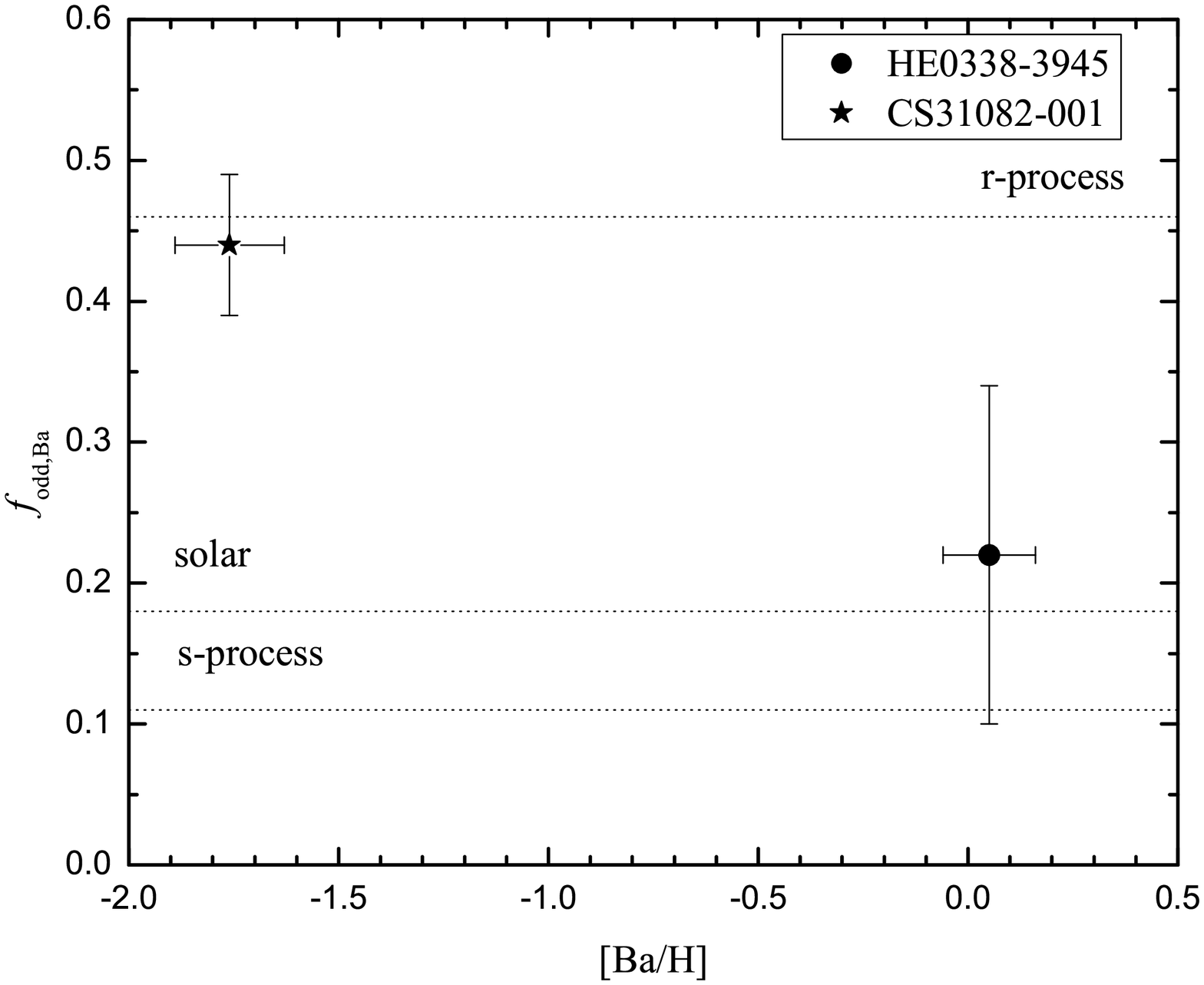}}
 \vspace{-8mm}
\resizebox{85mm}{!}{  \includegraphics{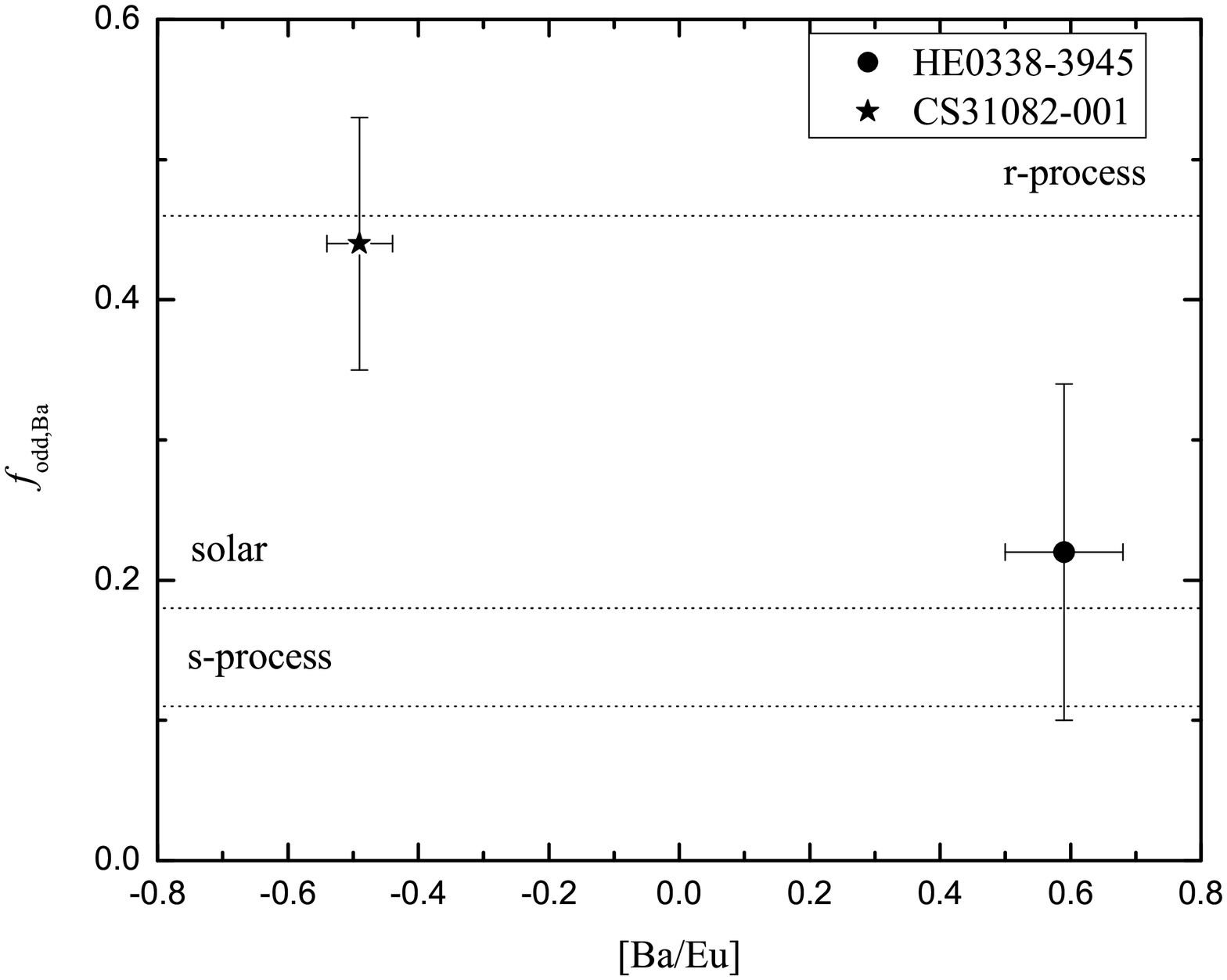}}
  \vspace{-5mm}
 \resizebox{85mm}{!}{ \includegraphics{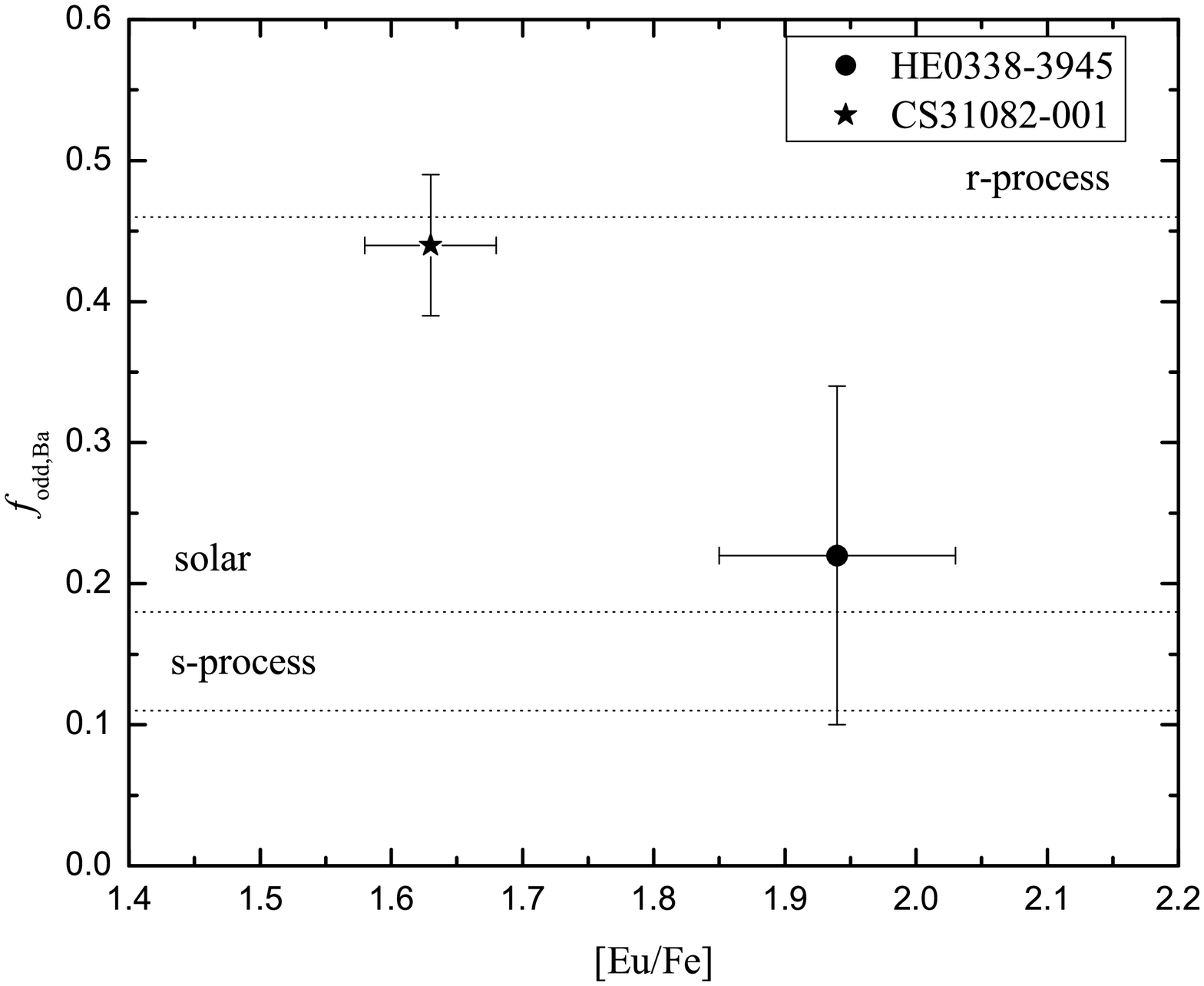}}
  \caption{The fraction of the odd Ba isotopes, \fodd, versus [Ba/H] (top panel), [Ba/Eu] (middle panel), 
  and [Eu/Fe] (bottom panel). Filled circle and Asterisk represent \he\ and \cs, respectively. 
  Uncertainties are shown by short horizontal and vertical lines. Dotted horizontal lines 
  indicate the \fodd\ values, 0.18 for the solar system, 0.46 for the pure r-process, and 0.11 for the pure s-process in production of the solar barium abundance predicted by \citet{arl99}. \label{foddBaH}}
\end{center}
\end{figure}

We present the \fodd\ values for \he\ and \cs\ as functions of [Ba/H], [Ba/Eu] and [Eu/Fe] in Figure~\ref{foddBaH}.
Where the values and the error bars of [Ba/H] for both stars are redetermined in this work, 
and those of [Eu/Fe] are adopted from \citet{jon06} and \citet{hil02}, respectively.

\subsection{CS\,31081-001}

\cs\ is a halo star with $\mathrm{[Fe/H]=-2.90\pm0.13}$, 
$\mathrm{[Ba/Fe]=1.14\pm0.05}$ (NLTE, obtained this work)
and $\mathrm{[Eu/Fe]=1.63\pm0.05}$ \citep{hil02}. As such, this star is known as an r-II star \citep{bee05}. 
Several groups \citep{hil02,sne08,cow11} have compared the abundance 
distribution of \cs\ with the solar r-process pattern, and found a good agreement,
especially for the elements heaver than barium. Based on this result, they suggested a pure-r-process origin 
for the heavier elements (i.e., $Z\geq56$) observed in \cs. 

From Figure~\ref{foddBaH}, we can see that 
\cs\ shows high value of \fodd, 0.43$\pm0.09$, and low abundance ratios of [Ba/H], -1.76$\pm0.13$, 
and [Ba/Eu], -0.49$\pm0.09$. 
The derived \fodd\ value of 0.43$\pm0.09$ is close to that of the pure r-process predicted by \citet{arl99}, which 
corresponds to a r-process contribution of about $91.4\pm25.7$\%
for barium. This implies that almost all of the barium in \cs\ is synthesized through a r-process.
The r-process contribution is calculated from the formula, r-process(\%) $= (f_\mathrm{odd,Ba}-0.11)/0.0035$, 
derived from \citet{gal10}. In addition, we note that \citet{aok03} have derived an average value of 0.44
for the isotope fraction of $^{151}$Eu for this star. \citet{arl99} expected that $94.2\%$ of europium in solar system 
material is originated from the r-process. Since europium has only two stable odd isotopes, ie. $^{151}$Eu and $^{153}$Eu,
the isotope fraction of $^{151}$Eu is often used to study the neutron-capture processes \citep[e.g.][]{sne02,aok03}. 
The $^{151}$Eu fraction is 0.541 for solar system europium originated from the pure-s-process, and 0.474 for that from
the pure-r-process \citep[both calculated from][]{arl99}. 
Obviously, the measured $^{151}$Eu fraction, 0.44, for \cs\ is close to the above value 0.474 calculated from the r-process
solar system residuals and 0.478 from the solar system material \citep{and89}.  

It can be seen that both the obtained isotope fractions of barium and europium
suggest that almost all of the heavy elements in \cs\ arise from a pure-r-process synthesis in 
SN\,II event or neutron star merger, as
barium and europium are usually regarded as the representative elements for the s- and r-process, respectively. 
This indicates that a universal pattern
for the r-process is supported not only by abundance pattern studies \citep{hil02,sne08,cow11},
but also by isotope level studies \citep[][and this work]{sne02,aok03}.

\subsection{\he}

\he\ is also a halo star with $\mathrm{[Fe/H]=-2.42\pm0.11}$, $\mathrm{[Ba/Fe]=2.47\pm0.11}$ 
(NLTE, obtained this work) and $\mathrm{[Eu/Fe]=1.94\pm0.09}$,
which belong to the so called CEMP-r/s star group \citep{bee05}. Here the values of [Fe/H]
and [Eu/Fe] are based on LTE results obtained by \citet{jon06}.
Its puzzling abundance 
pattern has been studied in detail by \citet{bis12} and \citet{cui10} based on AGB model 
calculations through fitting its abundance distribution. Based on the LTE abundance pattern of the heavy
elements, \citet{bis12} and \citet{cui10} predicted that the relative contributions of the 
r-process are about 3.4\% and 6.6\% for barium, respectively, for this object. 

From Figure~\ref{foddBaH}, it can be seen that \he\ has a low value of \fodd, $0.23\pm0.12$,
high abundance ratios of [Ba/H], 0.05$\pm0.11$, and [Ba/Eu], 0.59$\pm0.09$. 
The \fodd\ value of $0.23$ is slightly higher 
than both of the solar one, 0.18, and of the pure-s-process one predicted by \citet{arl99}, 0.11. 
The value of \fodd\,$=0.23\pm0.12$ corresponds to a r-process contribution of $34.3\pm34.3$\%
for barium. The r-process contribution is calculated from the same formula, 
r-process(\%) $= (f_\mathrm{odd,Ba}-0.11)/0.0035$, referred in Sec.5.1.
This indicates that the s-process still was the dominant neutron-capture process for the production of barium observed in \he, 
albeit its efficiency (65.7\%) is slightly lower than
that of the solar system. 
Obviously, 
the conclusions for the CEMP-r/s star \he\ based on AGB model calculations are also supported by  
isotope analysis. 

The bottom panel of Figure~\ref{foddBaH} shows an interesting fact that the CEMP-r/s star 
\he\ has a higher over-abundance of $\mathrm{[Eu/Fe]=1.94\pm0.09}$ than that of 
$\mathrm{[Eu/Fe]=1.63\pm0.05}$ for the r-II star \cs, which is the highest value obtained
up to now for r-II stars \citep[see][and reference therein]{mas10}. From the above analysis, both
the \fodd\ values and AGB model calculations indicate a s-process
domination of barium synthesis for \he. Both the AGB model calculations 
of \citet{cui10} and \citet{bis12} suggested that a pre-enrichment of the heavy elements from r-process 
leads to its high europium abundance. However, it is still a open question why the r-process shows
a higher efficiency on heavy element production for the CEMP-r/s star than that of
r-II stars. This topic exceeds this work, and will not be discussed detailedly here. 

\section{Conclusions}

In this paper, the results for the odd-isotope fractions of barium have been presented
for two metal-poor stars: the CEMP-r/s star \he\ and the r-II star \cs. This is the first time
the isotope fractions of barium in these two type stars have been measured, and these results will help us to
understand n-capture nucleosynthesis in low-metallicity environments at the isotope level. The measured
\fodd\ values are $0.23\pm0.12$ for \he\ 
corresponding to $34.3\pm34.3$\% of r-process contribution,
while $0.43\pm0.09$ for \cs corresponding to $91.4\pm25.7$\% of r-process contribution
to Ba, respectively.

The low \fodd\ value ($\sim0.23$) and thus the low r-process signature of barium in \he\ ($\sim34.3\%$) indicate that 
the heavy elements in this star formed through a mix of s-process and r-process synthesis, furthermore the s-process should be the dominant
neutron-process. This is consistent with studies based
on AGB model calculations for their abundance distributions \citep{cui10,bis12}. Since \he\ is a turn-off star 
 \citep[see Table~\ref{stellardata} and][]{jon06}, the s-process is not currently active. \citet{cui10} and \citet{bis12}
suggested that this star belongs to a binary system, and the enriched s-process material was polluted from its massive companion which have 
evolved through its AGB phase and now be an unseen white dwarf. They also suggested that the binary system formed from a cloud which 
has been polluted by a r-process event, such as SN\,II or neutron star merger. For \he, no difference in the barycentric radial velocities was found  
between the one measured from the snapshot spectrum taken on 15 October 2002 and that measured from the high quality 
spectrum taken in December 2002 \citep{jon06}.
Albeit the binarity of \he\ has not been confirmed up to now, 
the binarity of many CEMP-r/s stars has been confirmed in different works, e.g. for main-sequence stars CS\,29497-030 \citep{pre00}, 
CS\,29526-110 \citep{aok03b,tsa05}, HE\,2148-1247 by \citep{coh03}, for subgiant CS\,31062-050 \citep{aok03b,tsa05}, for giants
CS\,22948-027 \citep{pre01}, CS\,29497-034 \citep{pre01}, LP\,625-44 \citep{nor97,aok20}, HD\,209621 \citep{mcc90} 
and HE\,1405-0822 \citep{cui13b}. 
Recently, \citet{han16} reports their results for monitoring the radial velocities of 19 CEMP-s and three CEMP-r/s stars: 
15 CEMP-s and three CEMP-r/s stars belong to binary systems, which yields a binary frequency of 
$79\%$ for CEMP-s stars and 100\% for CEMP-r/s stars, respectively.

The high \fodd\ value ($\sim0.43$) for \cs\ means a r-process dominated regime for the barium origins in \cs. 
Also it suggests that almost all of the
heavy elements in \cs\ are produced by r-process, because
barium is usually regarded as a represent element of s-process, i.e. mainly produced
by s-process \citep{arl99}. This is consistent with the conclusions based on abundance pattern studies \citep{hil02,sne08,cow11}.
This result is also in line with the conclusions drawn from the isotope fraction for Eu
\citep{aok03}. This means that a universal pattern
for the r-process is supported not only by abundance pattern studies \citep{hil02,sne08,cow11},
but also by isotope level studies \citep[][and this work]{sne02,aok03}.

Obviously, the conclusions for the CEMP-r/s star, \he, and the r-II star, \cs, from the barium isotope analysis of this work,
agree well with the theoretical studies based on both their heavy element abundance patterns and/or their
europium isotope results.
Long-term radial-velocity monitoring is still desirable for \he.

\begin{acknowledgements}

 We heartly thank the anonymous referee for positive and constructive comments which helped to improve this paper greatly. This work is supported by  the National Natural Science Foundation of China under grants U1231119, 11321064, 11390371, 11273011, 
 11473033, 11547041, the China Postdoctoral Science Foundation under grant 2013M531587, the Natural Science Foundation of Hebei Province under grants A2011205102, A2014110008.

\end{acknowledgements}


\begin{thebibliography}{100}

\bibitem[Alonso et al.(1999)]{alo99} Alonso, A., Arribas, S., \& Mart$\mathrm{\acute{i}}$nez-Roger, C. 1999, A\&AS, 139, 335
\bibitem[Allen et al.(2012)]{all12} Allen, D. M., Ryan, S. G., Rossi, S., Beers, T. C., \& Tsangarides, S. A. 2012, \aap, 548, 34
\bibitem[Anders \& Grevesse(1989)]{and89} Anders, E., \& Grevesse, E. 1989, Geochimica et Cosmochimica Acta, 53, 197
\bibitem[Aoki et al.(2000)]{aok20} Aoki, W., Norris, J. E., Ryan, S. G., \& Beers, T. C., Ando, H., 2000, \apj, 536, 97
\bibitem[Aoki et al.(2003a)]{aok03} Aoki, W., Honda, S., Beers, T. C. \& Sneden, C. 2003, \apj, 586, 506
\bibitem[Aoki et al.(2003b)]{aok03b} Aoki, W. et al. 2003, \apj, 592, 67
\bibitem[Aoki et al.(2007)]{aok07} Aoki, W. et al. 2007, \apj,655, 492
\bibitem[Arlandini et al.(1999)]{arl99} Arlandini, C., K$\mathrm{\ddot{a}}$ppeler, F., Wisshak, K., et al. 1999, \apj, 525, 886
\bibitem[Asplund (2005)]{asp05} Asplund, M. 2005, \araa, 43, 481
\bibitem[Barbuy et al.(1997)]{bar97} Barbuy, B., Cayrel, C., Spite, M., et al. 1997, \aap, 317, L63
\bibitem[Barklem et al.(2005)]{bar05} Barklem, P. S., \& Christlieb, N., Beers, T. C., et al. 2005, \aap, 439, 129
\bibitem[Beers \& Christlieb(2005)]{bee05} Beers, T. C., Christlieb, N. 2005, \araa, 43, 531
\bibitem[Bistero et al.(2009)]{bis09} Bisterzo, S., Gallino, R., Straniero, O, \& Aoki, W. 2009, \pasa, 26, 314
\bibitem[Bistero et al.(2010)]{bis10} Bistero, S., Gallino, R., Straniero, O, Cristallo, S., \& K\"appeler, F. 2010, \mnras, 404, 1529
\bibitem[Bistero et al.(2012)]{bis12} Bistero, S., Gallino, R., Straniero, O, Cristallo, S., \& K\"appeler, F. 2012, \mnras, 422, 849
\bibitem[Burrows(2013)]{bur13} Burrows, A. 2013, Rev. Modern Phys., 85, 245
\bibitem[Busso el al.(1999)]{bus99}Busso, M., Gallino, R. \& Wasserburg, G. J. 1999, \araa, 37, 239
\bibitem[Cohen et al.(2003)]{coh03} Cohen, J. G., Christlieb, N., Qian, Y. Z., \& Wasserburg, G. J. 2003, \apj, 588,
1082
\bibitem[Collet et al.(2009)]{col09} Collet, R., Asplund, M., \& Nissen, P. E. 2009, \pasa, 26, 330
\bibitem[Cowan et al.(2011)]{cow11} Cowan, J. J., Roederer, I. U., Sneden, C., \& Lawler, J. E. 2011, in RR Lyrae
         Stars, Metal-Poor Stars, and the Galaxy, ed. A. McWilliam, 223
\bibitem[Cowley \& Frey(1989)]{cow89} Cowley, C. R., \& Frey, M. 1989, \apj, 346, 1030
\bibitem[Cui et al.(2010)]{cui10} Cui, W. Y., Zhang, J., Zhu, Z. Z., \& Zhang, B. 2010, \apj, 708, 51
\bibitem[Cui et al.(2013a)]{cui13a} Cui, W. Y. et al. 2013, Ap\&SS., 346, 477
\bibitem[Cui et al.(2013b)]{cui13b} Cui, W. Y., Sivarani, T., \& Christlieb, N. 2013, \aap, 558, 36
\bibitem[Cui et al.(2014)]{cui14} Cui, W. Y., Zhang, B., Zhao, G. 2014, Science China Physics, Mechanics \& Astronomy, 57, 1201
\bibitem[Gallagher et al.(2010)]{gal10} Gallagher, A. J., Ryan, S. G., Garc$\mathrm{\acute{i}}$a P$\mathrm{\acute{e}}$rez, A. E., \& Aoki, W. 2010, \aap, 523, A24
\bibitem[Gallagher et al.(2012)]{gal12} Gallagher, A. J., Ryan, S. G., Hosford, A., et al. 2012, \aap, 538, A118
\bibitem[Gallagher et al.(2015)]{gal15} Gallagher, A. J., Ludwig, H. G., Ryan, S. G., \& Aoki, W.  2015, \aap, 579, A94
\bibitem[Gallino et al.(1998)]{gal98} Gallino, R., Arlandini, C., Busso, M., et al. 1998, \apj, 497, 388
\bibitem[Gallino et al.(1999)]{gal99} Gallino, R., Busso, M., Lugaro, M., Travaglio, C., Arlandini, C., Vaglio, P.: In: Prantzos, N., Harissopulos, S. (eds.) Nuclei in the Cosmos V, p. 216. Fronti$\mathrm{\grave{e}}$res, Paris (1999)
\bibitem[Goriely et al.(2015)]{gor15} Goriely, S., Bauswein, A., Just, O., Pllumbi, E., \& Janka, H. T. 2015, \mnras, 452, 3894
\bibitem[Grupp(2004)]{gru04} Grupp, F. 2004, \aap, 420, 289
\bibitem[Grupp et al. (2009)]{gru09} Grupp, F., Kurucz, R. L., \& Tan, K. 2009, \aap, 503, 177
\bibitem[Hansen et al. (2016)]{han16}  Hansen, T. T., et al. 2016, \aap, 588, A3
\bibitem[Herwig(2005)]{her05} Herwig, F. 2005, \araa, 43, 435
\bibitem[Hill et al.(2000)]{hil00} Hill, V., Barbuy, B., Spite, M., et al. 2000, \aap, 353, 557
\bibitem[Hill et al.(2002)]{hil02} Hill, V., Plez, B., Cayrel, R., et al. 2002, \aap, 387, 560
\bibitem[Janka(2012)]{jan12} Janka, H.-T. 2012, Annu. Rev. Nucl. Part. Sci., 62, 407
\bibitem[Jonsell et al.(2006)]{jon06} Jonsell, K., Barklem, P. S., Gustafsson, B., et al., 2006, \aap, 451, 651
\bibitem[Karakas \& Lattanzio(2007)]{kar07} Karakas, A., \& Lattanzio, J. C. 2007, \pasa, 24, 103
\bibitem[Lambert \& Allende Prieto(2002)]{lam02} Lambert, D. L., \& Allende Prieto, C. 2002, \mnras, 335, 325
\bibitem[Lucatello et al.(2005)]{luc05} Lucatello, S., Tsangarides, S., Beers, T. C., et al. 2005, \apj, 625, 825
\bibitem[Lugaro et al.(2009)]{lug09} Lugaro, M., Campbell, S. W., \& de Mink, S. E. 2009, \pasa, 26, 322
\bibitem[Lugaro et al.(2012)]{lug12} Lugaro, M., Karakas, A. I., Stancliffe, R. J., \& Rijs, C. 2012, \apj, 747, 2
\bibitem[Magain \& Zhao(1993)]{mag93} Magain, P., \& Zhao, G. 1993, \aap, 268, L27
\bibitem[Magain(1995)]{mag95} Magain, P. 1995, \aap, 297, 686
\bibitem[Mashonkina et al.(1999)]{mas99} Mashonkina, L., Gehren, T. \& Bikmaev, I. 1999, \aap, 343, 519
\bibitem[Mashonkina \& Gehren(2001)]{mas01} Mashonkina, L., \& Gehren, T. 2001, \aap, 376, 232
\bibitem[Mashonkina \& Zhao(2006)]{mas06} Mashonkina, L., \& Zhao, G. 2006, \aap, 456, 313
\bibitem[Masseron et al.(2010)]{mas10} Masseron, T., et al. 2010, \aap, 509, A93
\bibitem[Mathews et al.(1992)]{mat92} Mathews, G.J., Bazan, G., Cowan, J.J. 1992, \apj, 391, 719
\bibitem[McClure \& Woodsworth(1990)]{mcc90} McClure, R. D., \& Woodsworth, A. W., 1990, \apj, 352, 709
\bibitem[Norris et al.(1997)]{nor97} Norris, J. E., Ryan, G., \& Beers, T. C., 1997, \apj, 488, 350
\bibitem[Pfeiffer et al.(1998)]{pfe98} Pfeiffer, M.J., Frank, C., Baumueller, D., Fuhrmann, K., \&
Gehren, T. 1998, A\&AS, 130, 381
\bibitem[Preston \& Sneden(2000)]{pre00} Preston, G. W., \& Sneden, C., 2000, \aj, 120, 1014
\bibitem[Preston \& Sneden(2001)]{pre01} Preston G. W., \& Sneden C., 2001, \aj, 122, 1545
\bibitem[Reetz(1991)]{ree91} Reetz, J. K. 1991, Diploma thesis, Universit$\mathrm{\ddot{a}}$t M$\mathrm{\ddot{u}}$nchen
\bibitem[Rosswog et al.(2000)]{ros00} Rosswog, S., Davies, M. B., Thielemann, F. K., Piran, T. 2000, \aap, 360, 171
\bibitem[Smith et al.(1998)]{smi98} Smith, V. V., Lambert, D. L., \& Nissen P. E. 1998, \apj, 506: 405
\bibitem[Sneden et al.(2002)]{sne02} Sneden, C., Cowan, J. J., Lawler, J. E., et al. 2002, \apj, 566, L25
\bibitem[Sneden et al.(2003)]{sne03} Sneden, C., Preston, G. W., \& Cowan, J. J. 2003, \apj, 592, 504
\bibitem[Sneden et al.(2008)]{sne08} Sneden, C., Cowan, J. J., \& Gallino, R. 2008 \araa, 46, 241
\bibitem[Short \& Hauschildt(2006)]{sho06} Short, C. I. \& Hauschildt, P. H. 2006, \apj, 641, 494
\bibitem[Spite et al.(2013)]{spi13} Spite, M. et al. 2013, \aap, 552, A107.
\bibitem[Takahashi et al.(1994)]{tak94} Takahashi, K., Witti, J., \& Janka, T. 1994, \aap, 286, 857
\bibitem[Terasawa et al.(2002)]{ter02} Terasawa, M., Sumiyoshi, K., Yamada, S., Suzuki, H., \& Kajino, T. 2002, \apj, 578, 137
\bibitem[Thompson(2003)]{tho03} Thompson, T. A. 2003, \apj, 585, L33
\bibitem[Travaglio et al.(1999)]{tra99} Travaglio, C., Galli, D., Gallino, R., et al. 1999, \apj, 521, 691
\bibitem[Truan(1981)]{tru81} Truan, J. W. 1981, \aap, 97, 391
\bibitem[Tsangarides(2005)]{tsa05} Tsangarides, S. A., 2005, PhD thesis, Open University (UK)
\bibitem[Wanajo et al.(2002)] {wan02} Wanajo, S., Itoh, N., Ishimaru, Y., Nozawa, S., \& Beers, T. C. 2002, \apj,  577, 853
\bibitem[Wanajo \& Ishimaru(2006)]{wan06} Wanajo, S., \& Ishimaru, Y. 2006, Nucl. Phys. A, 777, 676
\bibitem[Wanajo et al.(2011)]{wan11} Wanajo, S., Janka, H.-T., \& Mller, B. 2011, \apj, 726, L15
\bibitem[Wanajo et al.(2014)]{wan14} Wanajo, S. et al. 2014, \apj, 789, L39.
\bibitem[Woosley et al.(1994)]{woo94} Woosley, S. E., Wilson, J. R., Mathews, G. J., Hoffman, R. D., Meyer,
     B. S. 1994, \apj, 433, 229

\end{thebibliography}
\end{document}